\documentclass[10pt,a4paper]{article}

\usepackage{graphicx}
\usepackage{dcolumn}
\usepackage{bm}
\usepackage{authblk}
\usepackage{ragged2e}
\usepackage{parskip}
\usepackage{multicol}
\usepackage[usenames,dvipsnames]{color}
\definecolor{mygreen}{RGB}{31,115,110}
\usepackage[colorlinks=true,linkcolor=cyan,citecolor=cyan]{hyperref}%
\usepackage{soul}
\setstcolor{magenta}

\usepackage[sort&compress]{natbib}

\setcitestyle{comma,super,open={},close={}}

\usepackage{caption}
\usepackage{changepage}
\usepackage{marginnote}

\usepackage{tcolorbox}

\usepackage{array}
\newcolumntype{C}[1]{>{\centering\arraybackslash}p{#1}}

\usepackage[explicit]{titlesec}
\titleformat{\section}{\raggedright\bfseries\normalsize}{\arabic{section}.}{0pt}{{\MakeUppercase{#1}\vspace{.05ex}}}
\titlespacing{\section}{0pt}{14pt}{4pt}
\titleformat{\subsection}{\raggedright\bfseries\normalsize}{\arabic{subsection}.}{0pt}{{#1}}
\titlespacing{\section}{0pt}{14pt}{4pt}
\titleformat{\subsubsection}{\raggedright\itshape}{\arabic{subsubsection}.}{0pt}{}
\titlespacing{\subsubsection}{0pt}{16pt}{0pt}

\usepackage{geometry} 
\begin{document}


\title{\raggedright{\large{Perspective}}\\ \raggedright{\huge{Emerging ultra-wide band gap semiconductors for future high-frequency electronics}}}

\author[1]{\raggedright{Emily M. Garrity}}
\author[1]{Theodora Ciobanu\footnote{Current affiliation: University of Notre Dame}}
\author[2]{Andriy Zakutayev}
\author[1]{Vladan Stevanovi\'{c}\thanks{Correspondence: vstevano@mines.edu}}

\affil[1]{\raggedright{\small{Department of Materials \& Metallurgical Engineering, Colorado School of Mines, Golden, Colorado, USA}}}
\affil[2]{\raggedright{\small{Materials Science Center, National Renewable Energy Laboratory (NREL), Golden, Colorado, USA}}}

\maketitle

\begin{multicols}{3}
 
\section*{SUMMARY}
%
To meet the growing demands of advanced electronic systems, next-generation power and RF semiconductor devices must operate efficiently at higher power levels and switching frequencies while remaining compact. Current state-of-the-art GaN semiconductor devices alone cannot meet all these demands. Emerging ultra-wide band gap (UWBG) alternatives like diamond, BN, AlN, and Ga$_2$O$_3$, face significant challenges including limited wafer availability, doping difficulties, and thermal management constraints. Herein we conduct a high-throughput computational screening for new semiconductors for high-frequency electronics. In our analysis we compute the modeled Johnson and Baliga high-frequency figures of merit in combination with thermal conductivity to assess their potential for RF and power devices. We show that there are plenty of alternative materials to explore and conclude by discussing dopability and synthesis of select candidate materials. This study lays the foundation for discovering new semiconductors that can push the boundaries of performance in applications ranging from EV chargers and solid-state transformers to sub-THz communications and advanced radar technologies.

\section*{INTRODUCTION} 
%
The next generation of electronics demands devices capable of operating at both high frequencies and high power levels to enable breakthroughs across energy, mobility, and communications technologies. Radio frequency (RF) systems are advancing into the sub-THz frequency range to meet the performance demands of 6G communications, advanced radar, and next-generation satellite systems.\cite{tsao_ultrawide-bandgap_2018, hoo_teo_emerging_2021, cahoon_6g_2022} While higher frequencies allow faster data transfer and greater capacity, they also increase signal losses, requiring RF amplifiers with higher output power to maintain range and signal quality. These systems often rely on complementary power devices, such as high-frequency DC-boost converters used in vehicular communication systems.\cite{choi_fully_2016} More broadly, power electronics are seeing a growing demand for efficient, high-power conversion (AC$\leftrightarrow$DC) above 500 kHz, driven by emerging applications like fast-charging electric vehicles and long-range wireless power transfer.\cite{huang_power_2017, yuan_opportunities_2021, keshmiri_current_2020, choi_evaluation_2016} As both power and frequency demands rise, the development of advanced materials becomes central to enabling the electronics of the future.

\begin{figure*}
    \includegraphics[width=\textwidth]{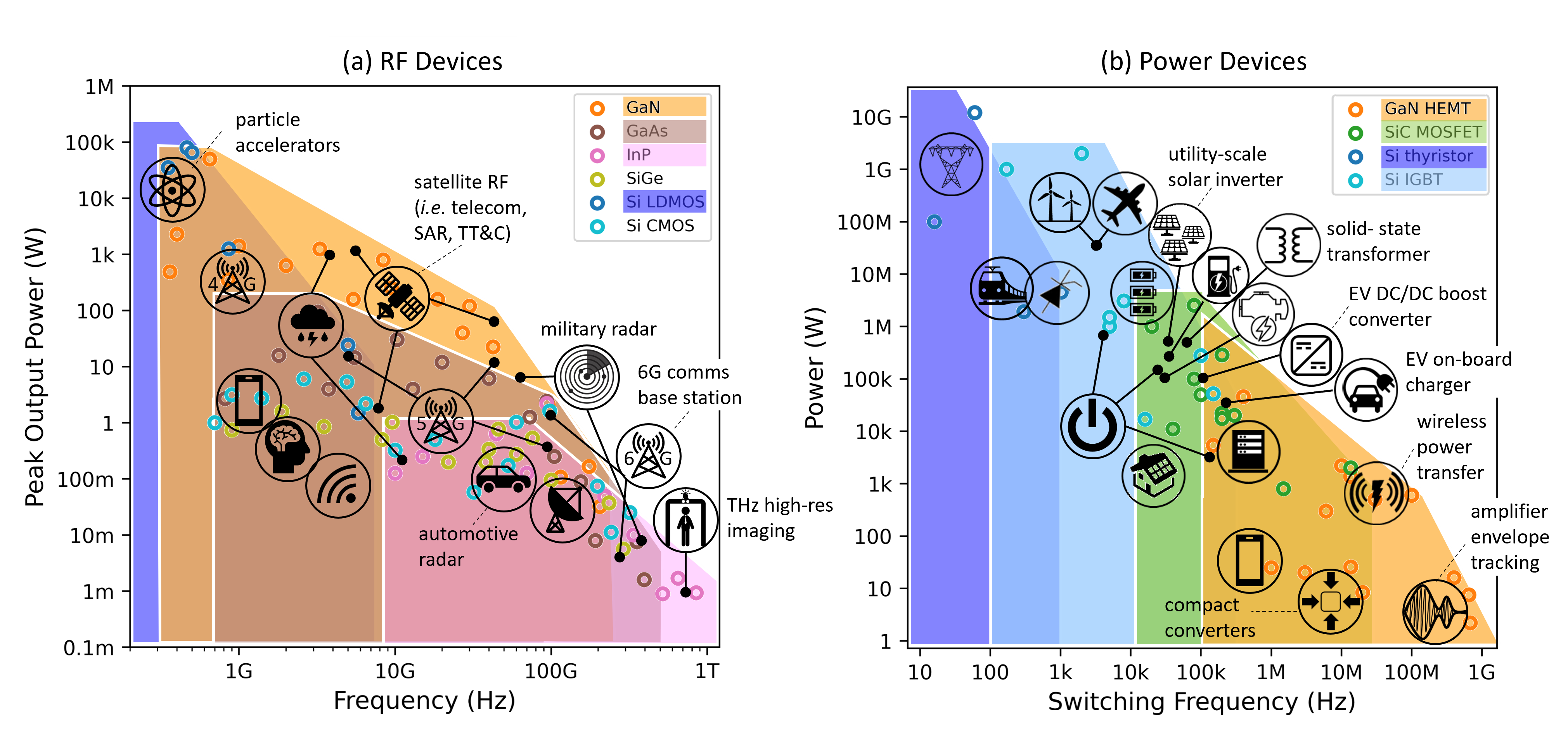}
    \caption{\textbf{State-of-the-art semiconductors and applications}\\
    (a) Peak output power and frequency ratings for commercialized RF power amplifier devices using Si, GaN, GaAs, SiGe, and InP. See Table S1 for application details. (b) Power and switching frequency ratings for commercialized power converting devices using Si, SiC, and GaN semiconductors. See Table S1 for application details. \\
    Dots are exemplar device ratings and symbols represent applications}
    \label{fig:1}
\end{figure*}

While narrow band gap semiconductors (E$_\mathrm{g}\leq$ 2 eV) still dominate today's power and RF electronics markets, they are increasingly limited by fundamental performance constraints (see Figure \ref{fig:1}). In RF, GaAs maintains a 38\% market share as the first commercialized compound semiconductor, driven by use in smartphones and satellites.\cite{SMR_rf_2024} Si CMOS and SiGe remain cost-effective solutions for high-volume RF applications like WiFi and cellular communications, despite lower frequency and power performance.\cite{solomko_fully_2009, steyaert_low-voltage_2002, zhang_time-mode-modulation_2024} InP is used in specialized cases above 100 GHz such as aerospace radar and medical instrumentation,\cite{urteaga_inp_2017} though its market growth is limited by power handling challenges. \cite{yole_status_2024} In power electronics, silicon holds over 90\% of the power market due to low cost and ease of fabrication.\cite{rosina_status_2023}

Adopting wider band gap semiconductors in power and RF electronics will allow higher operating ratings and result in massive energy savings due to enhanced efficiency. For instance, replacing all silicon inverters in photovoltaics with wide band gap devices (across a projected 75 TW of global capacity by 2050) could save 3,500 TWh annually, \cite{spejo_energy_2023-1} enough to power all of the homes in the U.S. three times over. Wide-gap materials also allow more compact and performant designs; for example, GaN power electronics in fast-charging smartphone adapters enable three times faster charging in devices that are half the size of their silicon equivalents.\cite{rafin_power_2023} 

\section*{State-of-the-art wide-gap semiconductors}
%
Gallium nitride (GaN) is the leading wide band gap semiconductor for high-frequency power and RF electronics. With a wide band gap (E$_\mathrm{g}$=3.4 eV) and high electron mobility ($\mu_n>$1200 cm$^2/$Vs), GaN supports faster switching, higher power density, and greater efficiency than narrow-gap materials. When GaN is paired with AlGaN alloy in a heterostructure, a lateral high-electron mobility transistor (HEMT) is formed which features a 2D electron gas channel with almost double the mobility of bulk GaN. The short gate length, reduced internal capacitance, and enhanced carrier transport properties enable frequencies above 100 MHz within a compact, light-weight package. While SiC remains dominant in high-voltage applications which require vertical devices, GaN’s lateral architecture makes it ideal for high-frequency inverters and amplifiers. As such, the power GaN market is rapidly expanding in the automotive (e.g., electric vehicle DC/DC converters, onboard chargers, LIDAR) and data center (e.g. power supplies) sectors. RF GaN, originally developed for defense, is now essential in 5G telecom, satellite communications, and radar systems and is projected to grow to \$2.07 billion by 2029 with a 12.8\% CAGR.\cite{rosina_status_2023}

Although GaN is the state-of-the-art material for high-frequency power conversion and amplification, it is approaching its theoretical performance limits. High-frequency GaN switches suffer from increased losses at higher powers, limiting their ability to deliver the multi-kilowatt power needed for some emerging applications shown in Figure \ref{fig:1}(b) such as wireless power transfer applicable to EVs or mid-air drone charging.\cite{jiang_high-efficiency_2019,keshmiri_current_2020} In addition, increasing operating limits results in a drop in efficiency. For example in a DC-boost converter, conversion efficiency can be close to 95\% at 500 kHz, but drops to 35\% above 600 MHz for the same output power (2-5 Watts) \cite{choi_evaluation_2016}. In RF applications, GaN HEMTs face similar scaling challenges. As frequencies rise above 30–100 GHz, GaN transistor performance is limited by size scaling effects set by material constraints such as carrier transit time and electric breakdown strength. For example, GaN power amplifiers with 0.25 $\mu$m gate lengths see power-added efficiency drop from 55\% at 1 GHz to below 25\% at 20 GHz \cite{sun_gan_2020}. These limitations restrict GaN’s future use in next-generation RF systems requiring both high frequency and high output power, such as those shown in Figure \ref{fig:1}(a) including 6G communications, long-range radar, cooperative sensing, and electronic warfare.

In addition to performance limitations, GaN faces manufacturing challenges, particularly related to creating high-power devices (above 10 kW). Achieving high voltages requires thick vertical transistors grown on bulk native substrates to minimize substrate-film mismatch defects, but bulk GaN is costly to grow and limited to small wafer sizes (2–4 inches) \cite{meneghini_gan-based_2021}. As a result, no commercial GaN devices exceed the kilovolt range. The highest demonstrated vertical GaN-on-GaN device is a 1.2-kV FinJFET \cite{liu_12_2020}. More commonly, GaN is used in lateral devices grown on silicon, sapphire, or SiC substrates, which are more cost-effective and scalable, but limited in their achievable power. While a 10-kV MC$^2$-HEMT of GaN has been demonstrated on 4-in sapphire wafers,\cite{xiao_multi-channel_2021} the output current of this device was just 0.06 ampere, limiting the output power to a few kW at most. 

Gallium also presents supply chain challenges. It is classified by the U.S. Department of Energy as an energy-critical material due to its reliance on by-product recovery from bauxite mining for primary production \cite{bauer_critical_2023}. To ensure a more resilient compound semiconductor supply chain, identifying alternative materials will be increasingly important.

\section*{Ultra-wide band gap alternatives}
%
Adopting ultra-wide band gap (UWBG) semiconductors (E$_\mathrm{g}>$ 3.4 eV) offers significant advantages for both power and RF electronics. For power electronics, their intrinsic properties enable higher operating voltages and switching frequencies while reducing conduction and switching losses. This translates to more compact, lightweight system designs due to reduced need for bulky cooling and passive components. In RF applications, UWBG materials extend the power-frequency limits beyond those of GaN, allowing for orders of magnitude improvements in power density at high frequencies.

\begin{figure*}
    \centering
    \includegraphics[width=0.93\textwidth]{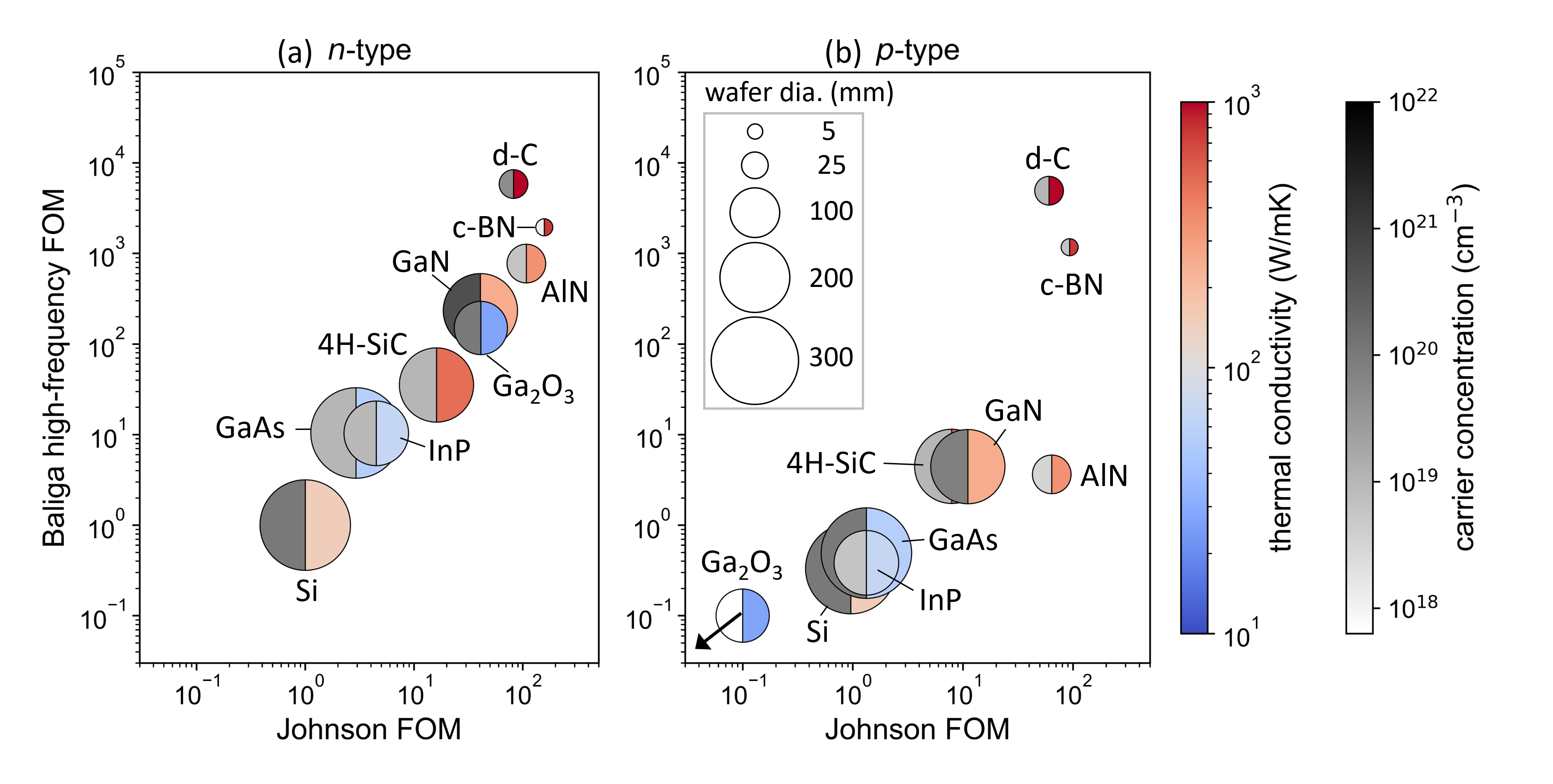}
    \caption{\textbf{Performance and wafer fabrication limits of established power and RF semiconductors}\\Theoretical operating limits and commercialization factors for (a) \textit{n}-type and (b) \textit{p}-type semiconductors used or in development for power and RF electronics. The Baliga high-frequency figure of merit is a measure of conduction and switching efficiency for power devices. The Johnson figure of merit is valid for high-frequency amplifiers. FOMs are calculated using best experimental data of bulk semiconductors (as available), which are listed in Table S3, and normalized to \textit{n}-type Si. The size of the circle indicates the maximum commercialized wafer diameter. Colors indicate the measured thermal conductivity and highest achieved carrier concentration through extrinsic doping.}
    \label{fig:2}
\end{figure*}

A handful of UWBG semiconductors have been considered for next-generation power and RF electronics, but challenges persist. Figure \ref{fig:2} illustrates the efficiency of various semiconductors in terms of their Johnson (JFOM) and Baliga High Frequency figures of merit (BHFFOM), relevant for low-power amplifiers and high-power fast-switching converters, respectively (see also Table S2). AlN, -Ga$_2$O$_3$, c-BN, and diamond offer orders of magnitude improvement in figures of merit compared to Si, SiC, and GaN. However, apart from Ga$_2$O$_3$, these alternative materials have not yet been grown as low-cost, low defect density single crystal substrates greater than 50 mm in diameter.\cite{tsao_ultrawide-bandgap_2018} Another challenge in device fabrication is achieving abrupt junctions with a low junction defect density ($<$10$^{12}$/cm$^{2}$) necessary for efficient carrier transport.\cite{tsao_ultrawide-bandgap_2018,setera_challenges_2022} Additionally, achieving controllable doping within the carrier concentration range of 10$^{15}$ to 10$^{20}$ cm$^{-3}$ remains challenging, particularly for AlN\cite{sarkar_n-_2018} and Ga$_2$O$_3$ (\textit{p}-type).\cite{zhang_recent_2020} Poor control of \textit{p}- or \textit{n}-type carriers can lead to low channel conduction and parasitic delays, even with optimized device design. The only UWBG candidate capable of growth by melt processes is $\beta$-Ga$_2$O$_3$. However, its low thermal conductivity limits use in the highest power and switching frequencies when heat dissipation is imperative.\cite{zhang_recent_2020} Therefore, we need new wide and ultrawide bandgap semiconductor options for advanced power and RF electronics.

\section*{Search for new semiconductors for future high-frequency electronics}
%
In our previous work we searched for alternatives to SiC for high power, high voltage applications by using high-throughput computations to rank over 1300 previously unexplored materials based on the well-known Baliga figure of merit (BFOM=$\varepsilon \mu E_b^3$) and thermal conductivity ($\kappa_L$). \cite{gorai_computational_2019,garrity_computational_2022-1} The Baliga figure of merit is a measure of the efficiency of on-state conduction and is relevant for vertical field effect transistors used in medium to high powers (kilowatts to tens of megawatts) and medium switching frequencies (ten kilohertz to several megahertz). 

Through this work, we identified a number of compounds that are competitive with 4H-SiC for high-power electronics. These include 49 oxides, 12 nitrides, 4 carbides, 2 sulfides, and 1 boride with predicted \textit{n}-type BFOM greater than 10 times that of Si and $\kappa_L$  greater than Ga$_2$O$_3$. The subsequent work includes a detailed evaluation of these top candidates to determine their dopability and feasibility for stable, quality thin-film growth. Among these proposed materials is UWBG InBO$_3$ (E$_g$=4.9-5.2 eV) whose predicted \textit{n}-type Baliga FOM is $>$6000 times that of Si. \cite{gorai_computational_2019, garrity_computational_2022-1} We predict that Zr is an effective dopant capable of providing $>10^{18}$ cm$^{-3}$ net electron concentrations and have grown thin films using pulsed laser deposition. \cite{garrity_defect_2023} 

While our previous material screening is encouraging for high-power devices, Figure \ref{fig:1} shows that there is a need for alternatives to GaN for high-switching-frequency power devices ($>>$MHz) and RF devices ($>>$GHz). Herein we expand upon our previous work, using DFT to evaluate both the \textit{n-} and \textit{p-}type Johnson and Baliga high-frequency figures of merit across the same dataset of 1300 compounds from the Inorganic Crystal Structures Database. Through this screening, we find plenty of alternatives for unipolar high-frequency electronics and suggest compounds to prioritize in future investigations.

\subsection*{Development of screening metrics}
The Johnson figure of merit (JFOM)\cite{johnson_physical_1965-1} defines the ultimate limit in the trade-off between voltage and switching frequency for lateral devices used in power amplifiers. Since frequency is inversely related to the carrier transit time from emitter to collector, it is maximized when the carriers have maximum velocity, which is typically at high electric fields ($>$ 10$^4$ volts/cm). Carrier transit time can be reduced further by shrinking the channel length, limited by the semiconductor's intrinsic breakdown field. Therefore JFOM can be written in terms of either operating limits or materials properties:
\begin{equation}
\mathrm{JFOM}= V_{m}f_t=\frac{E_b v_s}{2\pi},
\label{eq:JFOM}
\end{equation}
where $V_m$ is the maximum applied voltage, $f_t$ is the cutoff frequency, $E_b$ is the electric field at avalanche breakdown, and $v_s$ is the majority carrier saturation velocity. 

There is also a need for vertical power transistors that operate as GHz-frequency switches and require higher voltages than amplifiers. Therefore a different figure of merit, the Baliga high-frequency FOM,\cite{baliga_power_1989-2} is used for high-frequency power conversion applications. It is a measure of field-effect transistor efficiency, considering both on-state conduction losses and switching losses during input capacitor charging/discharging. The BHFFOM scales with material properties as:
\begin{equation}
    \mathrm{BHFFOM} \propto \mu E_b^2,
    \label{eq:BHFFOM}
\end{equation}
where $\mu$ is the electron or hole mobility. 

The material properties included in the figures of merit can be directly calculated via first principles, but this requires computing electron-phonon interactions and resulting scattering rates. These methods are not tractable for high-throughput evaluation. Instead, we use physics-motivated models for each relevant quantity which include descriptors that are easier to calculate for many materials. Models for the charge carrier mobility and intrinsic breakdown field were used successfully in our high-power screening \cite{gorai_computational_2019, garrity_computational_2022-1} and are adopted again in the high-frequency screening.  Calculation methods are detailed in supplemental materials.

\begin{figure*}
    \centering
    \includegraphics[width=0.95\textwidth]{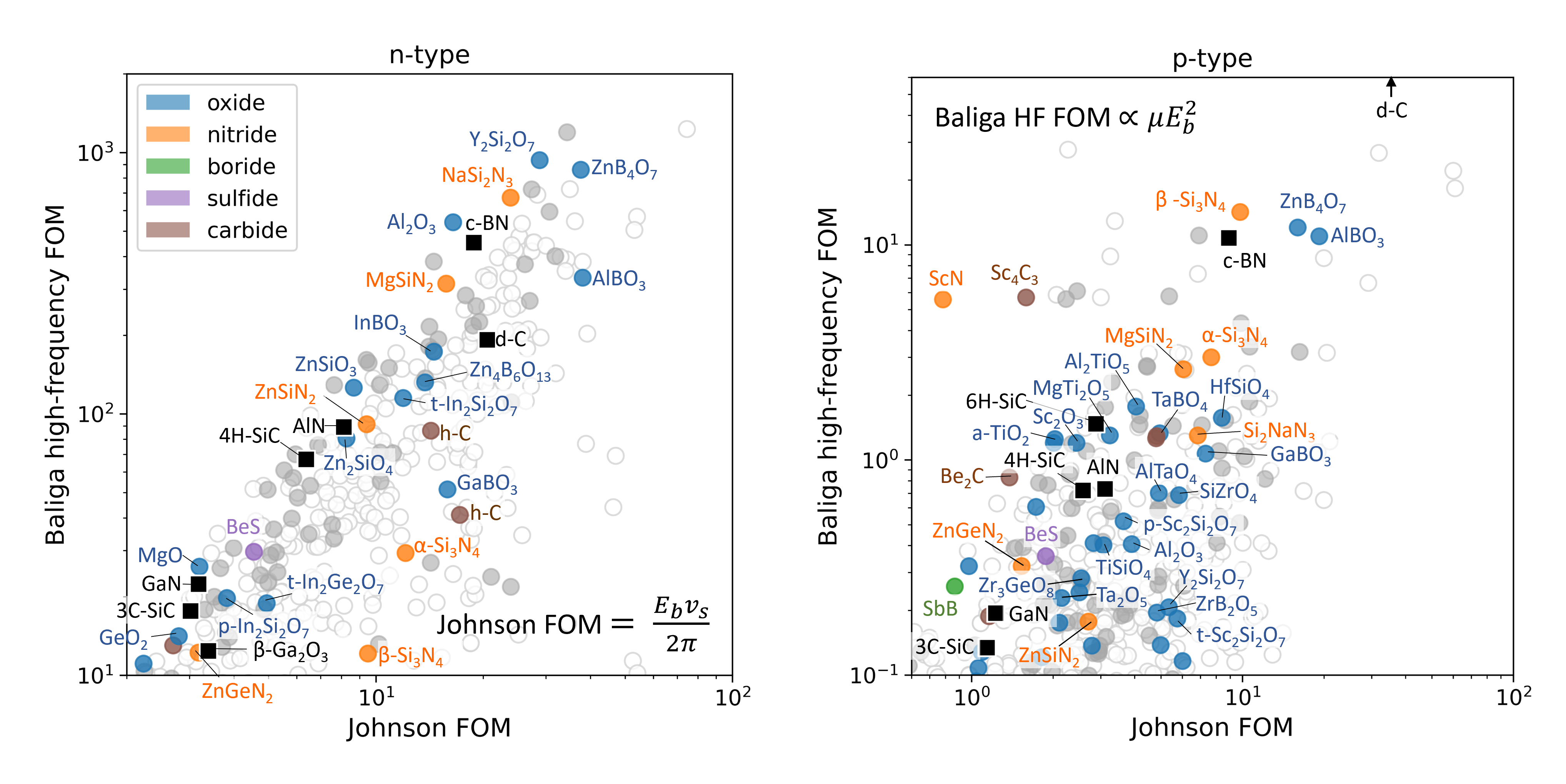}
    \caption{\textbf{Top candidates from the high-frequency semiconductor screening} \\Calculated (a) \textit{n}-type (electron conduction) and (b) \textit{p}-type (hole conduction)  high-frequency figures of merit for screened candidates. Plots are zoomed-in on largest FOMs. Data with high modeled thermal conductivity ($\kappa_L>25$ W/mK) are colored by anion. Data with modeled  $\kappa_L$ between 10-25 W/mK are dark gray and $\kappa_L<10$ W/mK are light gray. See also Tables S4 and S5.}
    \label{fig:3}
\end{figure*}

In addition, high-frequency applications require evaluation of the saturation velocity of charge carriers for which we use a textbook model:
\begin{equation}
v_s = \bigg(\frac{8 \hbar \omega_{0}}{3 \pi m^*_b}\bigg)^{1/2},
\label{eq:vsat}
\end{equation}
where $\hbar \omega_0$ is the energy of the optical phonon dominant in scattering and $m^*_b$ is the conduction (valence) band effective mass for the valley (pocket) where the free electrons (holes) reside. While multiple scattering mechanisms may contribute to velocity saturation, the most influential ones at high electric fields are related to high energy optical phonon emission. \cite{jacoboni_monte_1983} Thus, in eq.~\ref{eq:vsat} we use the highest phonon energy at $\Gamma$, assuming that optical modes have low dispersion. 

In addition, since eq.~\ref{eq:vsat} uses a single value for $m^*_b$, it assumes that scattering occurs within a single spherical charge carrier pocket. The assumption that electrons reside in the lowest lying conduction band, even at electric fields near 1000 kV/cm, is valid for most materials. However, materials like GaAs and group-III nitrides have additional heavier low-lying bands which electrons tend to scatter into under high electric fields. This causes carrier drift velocity to peak then saturate to a lower value as electric field is increased. Short-channel ($<$0.2 $\mathrm{\mu m}$) devices can take advantage of velocity overshoot, operating with an applied electric field that maximizes velocity during carrier transit time. This has been demonstrated in Al$_{x}$Ga$_{1-x}$N/GaN HEMTs where saturation velocities are twice that of bulk GaN, up to 3.1 $\times 10^7$ cm/s. \cite{barker_bulk_2005} By construction, eq.~\ref{eq:vsat} is a prediction of the peak velocity which we use in our screening rather than true saturation. Comparisons of measured and modeled (eq.~\ref{eq:vsat}) $v_s$ for electrons and holes of 14 common semiconductors is available in Figure S1, demonstrating satisfactory accuracy for screening purposes.

\section*{Results and Discussion}
In Figure \ref{fig:3} we plot the calculated figures of merit (normalized to \textit{n}-type Si) for our 1300 candidates alongside a few reference materials, zoomed-in on the top performers. Predicted BHFFOM spans eight orders of magnitude and JFOM spans three orders of magnitude across all calculated materials, indicating considerable differentiation of predicted performance. It should be noted that because of the known underestimated band gap of DFT, some differences in the ranking of reference materials of Figure \ref{fig:2} and \ref{fig:3} is to be expected. However, the relative FOM rankings generally follow the expected trends. Diamond and c-BN are among the highest ranked, followed by UWBG AlN, then wide-gap SiC and GaN, and finally narrow-gap Si. 

Interestingly, we observe that there are more than 250 candidates with FOMs that surpass or are near that of GaN for both \textit{n}-type and \textit{p}-type searches. Among the top \textit{n}-type candidates, the predicted electron saturation velocities are within 1.5-4 $\times 10^7$ cm/s, while the breakdown fields range from 1.5-32 MV/cm. Similarly, the top \textit{p}-type candidates have hole saturation velocities within one order of magnitude while the breakdown field varies by close to two orders of magnitude. This suggests that the breakdown field has more influence in differentiation than saturation velocity among the top candidates.

\subsection*{Considering thermal conductivity}
A known restriction of the figures of merit is the exclusion of thermal performance evaluation. The thermal limit of AlGaN/GaN HEMTs is estimated to be 100 W/mm for continuous power operation, constraining output power when operating below 50 GHz frequency. \cite{coffie_high_2020} To improve power dissipation, any replacement semiconductor should have large thermal conductivity. Thus we predict $\kappa_L$ as an additional screening metric, using a modified Debye-Callaway model as in Ref.~\cite{garrity_computational_2022-1, gorai_computational_2019} and detailed in supplemental materials.

In Figure \ref{fig:3}, we color those materials with predicted $\kappa_L>25$ W/m-K according to their anion. This $\kappa_L$ cutoff is larger than Ga$_2$O$_3$ and close to GaN predictions with some buffer left for model accuracy (about half order of magnitude). There are 20 \textit{n}-type candidates and 42 \textit{p}-type candidates with $\kappa_L>25$ W/m-K and FOMs greater than 85\% of GaN. 

Upon closer inspection, we see that the majority of the remaining candidates are oxides, which reflects the starting dataset. There are also four nitrides for \textit{n}-type: Si$_3$N$_4$ (space group 159), NaSi$_2$N$_3$, MgSiN$_2$, and ZnSiN$_2$, and an additional two for \textit{p}-type: Si$_3$N$_4$ (space group 176) and ZnGeN$_2$. Despite there being 300 sulfides in the starting dataset, only BeS appears among top candidate. Most others have low predicted FOMs and $\kappa_L$. For carbides, \textit{n}-type candidates are only hexagonal diamonds. \textit{P}-type carbides include GeC, Be$_2$C, B$_4$C, and Sc$_4$C$_3$. While over 28\% of the starting dataset are quaternaries, only two appear since the majority of these materials have low $\kappa_L$ of less than 15 W/m-K. The strongest remaining candidates are mostly ternaries, which are largely unexplored for electronics. 

\subsection*{Addressing additional semiconductor criteria}
To fully determine viability as high-frequency semiconductors, additional criteria must be addressed. These include doping, thermodynamic stability, thin film growth, and the availability of complementary semiconductors for heterostructures. Within the context of this paper, we discuss known information for the candidates. 

First priority is dopability as materials with larger band gaps may rank highly but are often not dopable. The most upper-right hand corner of Figure \ref{fig:3} includes the largest band gap candidates (calculated E$_g>6$ eV), some of which are known or predicted insulators. These include Al$_2$O$_3$, $\beta$-Si$_3$N$_4$, Sc$_2$O$_3$, (Al,Ga)BO$_3$, \cite{garrity_defect_2023} and Y$_2$Si$_2$O$_7$. \cite{gorai_computational_2019} Other materials previously studied as phosphors, including (Hf,Zr)SiO$_4$, (Al,B)TaO$_4$, Al$_2$TiO$_4$, Si$_2$NaN$_3$, Zn-B-O ternaries, and $\alpha$-Si$_3$N$_4$, are likely insulators but have not been explicitly reviewed for semiconductor doping. 

There are promising alternatives that we already know are dopable. Among \textit{n}-type candidates, InBO$_3$ and thortveitite t-In$_2$(Si,Ge)$_2$O$_7$ also appear in our previous search for high-power semiconductors and are predicted to be extrinsically dopable. Zirconium doped InBO$_3$ has predicted net donor concentrations up to $>10^{18}$ cm$^{-3}$ and has been succesfully grown as a thin film via pulsed laser deposition. \cite{garrity_computational_2022-1, garrity_defect_2023} For In$_2$Ge$_2$O$_7$, Zr-doped polycrystalline thin films had measured electron concentrations of 10$^{14}$–10$^{16}$ cm$^{-3}$ even under O-rich conditions, while theoretical predictions suggest single crystals grown under optimal \textit{n}-type conditions could reach up to 10$^{19}$ cm$^{-3}$ electrons. \cite{lee_stability_2025} ZnGeN$_2$, a member of a family of (Mg,Zn)(Si,Ge)N$_2$ orthorhombic nitrides studied for optoelectronics, is \textit{n}-type dopable as well. Thin films of unintentionally-doped and phosphorous-doped ZnGeN$_2$ have achieved up to 10$^{19}$ cm$^{-3}$ net electrons. \cite{adamski_optimizing_2019, martinez_synthesis_2017} Other members of this family, like MgSiN$_2$ and ZnSiN$_2$, suffer from compensation.

If controllable doping can be realized, the ternary nitrides are prime candidates for use in HEMTs due to their strong polarization and close lattice-matching to III-N. A proposed heterojunction of ZnSiN$_2$/AlN is expected to greatly outperform AlGaN/GaN HEMTs. \cite{adamski_band_2020} Similarly, heterostructures and alloying among In$_2$(Si,Ge)$_2$O$_7$ and (In,Ga,Al)BO$_3$ could open up the possibilities for band gap engineering and the creation of HEMT devices.     

Lastly, if a target application does not require as high thermal conductivity, then we can lower the $\kappa_L$ limit to 10 W/m-K and consider an additional 46 \textit{n}-type and 57 \textit{p}-type candidates, including MgGeN$_2$, as depicted by gray markers in Figure \ref{fig:3} and listed in Tables S4 and S5.

\section*{Outlook} In conclusion, the results of our high-throughput screening are promising, indicating that there are numerous alternatives to GaN worth exploring for high-frequency power and RF electronics. Of course, this is just a first step and additional studies are needed to know which of these candidates will realize their full potential. Among those identified, we find materials that are suitable for both high power and high switching frequency and for which the dopability has already been predicted: InBO$_3$, In$_2$Si$_2$O$_7$, In$_2$Ge$_2$O$_7$, and ZnGeN$_2$. For the rest of the candidates, subsequent studies should, in our view, begin with a dopability assessment followed by thin film growth, and, depending on the device architecture, identification of suitable contacts and complementary materials for heterostructures. The realization of devices based on these new semiconductors could enable faster, more compact EV chargers, unlock high-power wireless power transfer, and extend the capabilities of RF systems for next-generation communications, radar, and sensing technologies.   

\section*{ACKNOWLEDGEMENTS}
%
This work was authored in part at the National Renewable Energy Laboratory (NREL) for the U.S. Department of Energy (DOE) under Contract No. DEAC36-08GO28308. Funding provided as part of APEX (A Center for Power Electronics Materials and Manufacturing Exploration), an Energy Frontier Research Center funded by the U.S. Department of Energy, Office of Science, Basic Energy Sciences under Award \#ERW0345 (application and performance analysis); and by the Laboratory Directed Research and Development (LDRD) program at NREL (materials figure of merit calculations). The research was performed using computational resources sponsored by the DOE’s Office of Energy Efficiency and Renewable Energy located at NREL. The authors are grateful to Prof. Darrell Schlom for the insightful discussions which helped shape this research. The views expressed in the article do not necessarily represent the views of the DOE or the U.S. Government.

\section*{Author contributions}
Conceptualization, E.G., A.Z., and V.S.;
Computations, E.G.;
Analysis, E.G., T.C., and V.S.;
Writing and editing, E.G., A.Z., and V.S.

\section*{declaration of interests}
The authors declare no competing interests.

\section*{SUPPLEMENTAL INFORMATION}
Document S1. Tables S1-S5, Figure S1, and calculation methods for material property models. 


\end{multicols}
\end{document}